# Controlling the harmonic conversion efficiency in semiconductor superlattices by interface roughness design

Apostolos Apostolakis[a)], and Mauro F. Pereira[b)]

*Department of Condensed Matter Theory, Institute of Physics CAS, Na Slovance 1999/2, 182 21 Prague, Czech Republic*

*In semiconductor superlattices, when Bragg oscillating electrons interact with an input electromagnetic field, frequency multiplication is possible. An ideal superlattice has a purely antisymmetric voltage current response and can thus produce only odd harmonics. However, real world superlattices can also have even harmonic response and that increases the range of possible output frequencies. These effects have been recently explained with a predictive model that combines an Ansatz solution for the Boltzmann Equation with a Nonequilibrium Green's Functions approach. This predictive tool, coupled with recent progress on GHz input sources, support the growing interest in developing compact room temperature devices that can operate from the GHz to the THz range. The natural question to ask is what efficiencies can be expected. This paper addresses this issue by investigating power-conversion efficiency in irradiated semiconductor superlattices. Interface imperfections are consistently included in the theory and they strongly influence the power output of both odd and even harmonics. Good agreement is obtained for predicted odd harmonic outputs with experimental data for a wide frequency range. The intrinsic conversion efficiency used is based on the estimated amplitude of the input field inside the sample and thus independent of geometrical factors that characterize different setups. The method opens the possibility of designing even harmonic output power by controlling the interface quality.*

High power coherent sources for the whole Gigahertz (GHz)-Terahertz (THz) - Mid Infrared (MIR) ranges, operating at room temperature are in demand for a myriad of applications and nonlinear effects are evolving into the dominant solutions. Difference frequency generation via resonant optical nonlinearities pumped by a MIR quantum cascade laser [1] is encouraging for the THz-MIR. The combination of input from superlattice electronic devices (SLEDs) [2] with semiconductor superlattice (SSL) multipliers [3, 4] is highly promising since for the GHz-THz: (i) SLEDS have reached a record 4.2 mW power output in the fundamental mode at 145 GHz at room temperature [2]. (ii) Synchronization between SLLs leads to a dramatic increase in output power [5].

From a fundamental point of view, nonlinearities in SSLs provide numerous opportunities to create and develop spectroscopic schemes, including harmonic generation, mixing, detecting and parametric processes of high-frequency electromagnetic radiation [2-10]. In fact, negative differential conductivity (NDC) was early recognized by Esaki and Tsu as a key ingredient for the generation of harmonics of microwave and THz radiation [11, 12]. Higher-order multipliers based on SSL devices have already been demonstrated [3-10] and significant enhancement of the generated power has been observed at certain threshold amplitude of the input field and attributed either to nonlinearities induced by the domain formation [7] or to the onset of the parametric amplification [8]. However, discrepancies between simulations and experiments were interpreted as optical losses [7, 8] without a direct modelling that allows the calculation of conversion efficiency accurately. Recently, we suggested an alternative approach to describe controllable THz-GHz nonlinearities covering both "even" and "odd" nonlinear responses [3, 4]. We obtained good agreement between experimental and theoretical results for the output power of SSL under the influence of a GHz electric field with fixed parameters (input frequency ω, field amplitude $E_{ac}$). More interesting, however, was the spontaneous frequency multiplication effect for even harmonics in unbiased SSL [3]. We note that significant gain had been previously predicted at even harmonics [13-15] of an unbiased high-frequency electric field but have been attributed either to parametric amplification [13, 14] or other parametric effects which require the existence of an internal electric field in the device [15]. However, even harmonic generation due to the differences in the interface structure of the superlattice layers and therefore different interface roughness (elastic) scattering rates are mostly unexplored and are studied in this paper, even though it is known that this process plays an important role in electron transport in SSLs. As a matter of fact, previous experimental work has focused on trying to understand how the dephasing mechanisms of Bloch oscillations and the electron mobility are related specifically to (elastic) interface roughness scattering [16, 17]. Special attention has been given to nonlinear balance-equation transport dynamics [6, 18], which has been systematically employed to make prediction for different terahertz amplification and generation schemes [14, 19]. This approach—which follows from Boltzmann transport equation—allows one to take into account scattering processes which change both momentum and energy but it depends heavily on fitting the analytical equations to experimental data or assuming purely phenomenological parameters [20]. Furthermore, elastic scattering was included in these models by means of an ansatz which describes forward and reverse scattering with equal weights.

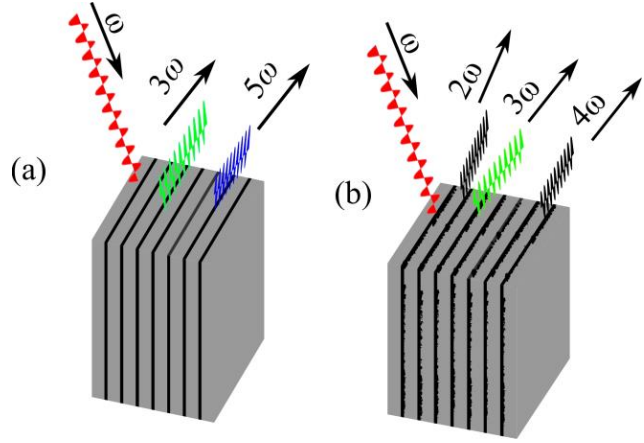

FIG. 1. Schematic representation of: (a) superlattice multiplier (SSLM) with symmetric interfaces leading to purely antisymmetric current-voltage and capable of generating only odd harmonics. (b) SSLM with nonsymmetric interfaces capable of generating both even and odd harmonics.

In this paper, we explore the fundamental limits to conversion efficiency expected for a SSL under the action of an oscillating electric field. Among the results presented here, we highlight: (i) Transport and power emission calculations under the additional influence of the asymmetry in current flow induced by interface roughness scattering. (ii) Harmonic power efficiency calculated directly from the Poynting vector, by estimating the field inside the sample with results consistent with measurements found in the literature for odd harmonics. (iii) Study of the predicted efficiency for even harmonics as a function of controlled symmetry breaking at the current flow level.

We are not interested in circuit-equivalent approaches which might be reliable for a quasi-classical solid state oscillator [9] but which are not predictive if spontaneous frequency multiplication effect takes place [3, 4]. Then, we adopt another way in which we describe an optical method to precisely determine the intrinsic conversion efficiency of the excited SL sample (see Fig. 1). Our predictions are based in an approach which does not need phenomenological parameter fitting, since our Nonequilibrium Green's Functions (NEGF) implementation can deliver input for an analytical Ansatz solution for the Boltzmann equation, including current flow asymmetry. We demonstrate that a semiconductor superlattice allows the conversion of input power to third-harmonic radiation with an efficiency of about 2 %. Our calculations are consistent with previous experimental studies and incorporate the possibility of imperfections in the structure (see Fig. 1.b). This latter characteristic is crucial since our microscopic approach allows to foresee theoretically the generation of the even harmonics, whereas in a high-quality superlattice only the odd harmonics would be present.

We are interested in the electron dynamics under the action of a time-dependent electric field, which consists of constant and oscillating parts, $E(t) = E_{dc} + E_{ac}\cos(\omega t)$. For this input field $E(t)$, which is parallel to the growth direction of the SSL with period $d$, the general current response can be written as

$$j(t) = j_{dc} + \sum_{l=-\infty}^{\infty} j_l^c \cos(l\omega t) + j_l^s \sin(l\omega t),$$

$$j_{dc} = \sum_{p=-\infty}^{\infty} J_p^2(\alpha) Y(U),$$

$$j_l^c = \sum_{p=-\infty}^{\infty} J_p(\alpha)[J_{p+l}(a) + J_{p-l}(a)] \, Y(U), \quad (1)$$

$$j_l^s = \sum_{p=-\infty}^{\infty} J_p(\alpha)[J_{p+l}(a) - J_{p-l}(a)] \, K(U).$$

Here $J_p$ is the Bessel function of the first kind and order , and $U = eE_{dc}d + p\hbar\omega$ is the resulting effective potential difference which electrons experience instead of the bare potential due to the dc bias [3, 4, 6]. If the distribution of electrons is approximately homogeneous, the local transport properties are governed by the global voltage-current characteristic of the device. The parameter $\alpha = eE_{ac}d/\hbar\omega$ which appears automatically as a consequence of our model controls the nonlinear response of the system and its strong deviation from typical N-order susceptibilities. It also sets the scale for dynamic-localization phenomena [21]. If the distribution of electrons is approximately homogeneous, the local transport properties are governed by the global voltage-current characteristic of the device. Throughout this discussion, we restrict ourselves to a homogeneous field distribution $E(t)$ for which the current is homogeneous over the superlattice direction and it is sufficiently well described by Eq. (1).

From Eq. (1), we can identify functions $Y$ and $K$ which hold for the miniband transport within the relaxation time $\tau$ approximation [3-4, 6],

$$Y(U) = j_0 \frac{2U/\Gamma}{1+(U/\Gamma)^2}, \quad K(U) = \frac{2j_0}{1+(U/\Gamma)^2}. \quad (2)$$

Here, $\Gamma = \hbar/\tau$ is the scattering induced broadening and $j_0 = e\frac{\Delta d}{2\hbar}/(2\pi)^3 \int d^3k \cos(k_x d) n_F(\mathbf{k})$ is the peak current $j_0$, corresponding to $U = U_c$ [$U_c \equiv \Gamma$]. In the Boltzmann equation approach [6, 11], the explicit formula for $j_0$ is determined by the fermi distribution $n_F(\mathbf{k})$ and the standard tight-binding dispersion relation $\varepsilon(k_x) = -\frac{\Delta}{2}\cos k_x d$, where $\varepsilon(k_x)$ is the electron energy, $k_x$ is the quasimomentum and $\Delta$ is the miniband width. In our hybrid approach we obtain $\Gamma$ and $j_0$ by direct comparison with the calculated static NEGF current-voltage curves [3, 4]. One can see from Eqs. (1) and (2) that in the presence of an ac field the voltage current characteristic is given by a sum of shifted Esaki-Tsu characteristics [11]. We proceed by considering that we are in steady state and no transient effects are present. Then, Eq. (1) implies that without an applied of background symmetry-breaking field, the application of an oscillating field to a SL characterized by a current $j(t)$ can produce odd harmonics for structures with $j_{dc}(E_{dc}) = -j_{dc}(-E_{dc})$. This remarkable symmetry property is a manifestation of a superlattice structure with symmetric interfaces, even though they may have roughness fluctuations. Nevertheless, it is not always realistic to assume that the interface of GaAs over $Al_{1-x}Ga_xAs$ demonstrates the same features as the one of $Al_{1-x}Ga_xAs$ over GaAs. Quite on the contrary, they are usually of different quality, which is revealed by an asymmetric current flow. In particular, from detailed comparison between theory and experiment, we showed that the aforementioned asymmetry introduces a variation to the traverse transport parallel to the SSL layers inherently dependent on the growth direction [3, 4]. Therefore, the NEFG formalism which was used to treat numerically the impact of different interfaces, revealed that the initial Esaki-Tsu approximation could be adapted according to [3, 4]

$$j_0 = \begin{cases} j_0^-, & U < 0 \\ j_0^+, & U \geq 0 \end{cases}, \quad \Gamma = \begin{cases} \Gamma^-, & U < 0 \\ \Gamma^+, & U \geq 0. \end{cases} \tag{3}$$

The main parameters extracted from the NEGF calculations (see Appendix) and used in this ansatz solution were: $\Gamma^+, \Gamma^- =21, 20$ meV and $j_0^+, j_0^- =2.14, 1.94 \times 10^9$ A/m$^2$. In this paper, to estimate the generated power and the power-conversion efficiency we will use the following parameters for a GaAs/AlGaAs SSL: period d=6.23 nm, electron density $N = 1.5 \times 10^{18}$ cm$^{-3}$ and refractive index $n_r = \sqrt{13}$ (GaAs). The calculated relaxation rate is $\tau = \hbar/\Gamma = 31$ fs, see Refs. [3, 4]. This is a typical SSL structure, which has a miniband width $\Delta \sim 140$ meV. In physical terms this nonsymmetrical generalization permits us to introduce an additional parameter $\delta = j_0^+/j_0^-$ (see Appendix) coined as asymmetry coefficient and determined by the equation

$$\frac{j_0^+}{\Gamma^+} = \frac{j_0^-}{\Gamma^-}. \tag{4}$$

When $\delta = 1$, the characteristic scattering time of miniband electrons is $\tau = 31$ fs. In order to describe the asymmetry in the simplest possible way, we fix the parameters $j_0^+, \Gamma^+$ and modify $j_0^-, \Gamma^-$, which characterize the inverse-polarity current. The resulting asymmetry parameter $\delta$ is varied within reasonable limits.

Previous studies, which however could only predict odd harmonics, in contrast to our more complete approach, have demonstrated that the frequency multiplication mechanism stems from possible the direct interaction of the input field with Bloch oscillating electrons [21]. The exact condition which determines the onset of Bloch oscillations in a SL for an unbiased oscillating field corresponds to the critical value of $a_c = U_c/h\nu$, where the input field amplitude ($E_{ac}$) equals the critical field $U_c/(ed)$ after which the static I-V shows NDC. This is also related to the relaxation time of the sample, since $\tau = \frac{\hbar}{\Gamma} = \frac{\hbar}{U_c}$.

From Electromagnetic Theory, the power emitted by the currents induced by the oscillating field is calculated from the Poynting vector [3, 4]. Therefore, the driving term for the power emitted by the l$^{th}$ harmonic is

$$I_l(\omega) = <j(t)\cos(l\omega t)>^2 + <j(t)\sin(l\omega t)>^2, \tag{5}$$

where $j(t)$ is the current density induced in the SSL by the total field $E(t)$ and the averaging $<\cdots>_t$ is performed over the period $T = 2\pi/\omega = 1/\nu$. The corresponding generated power [3, 4] is related to $I_l$ to by

$$P_l(\omega) = \frac{A \mu_0 c L^2}{8 n_r} I_l(\omega), \tag{6}$$

where waveguide effects have been neglected. Here $A$ is the area of the mesa of a superlattice element, $\mu_0$ and $c$ are the permeability and speed of light in the free space, $L$ is the effective path length through the crystal, $n_r$ is the refractive index of the SL material.

Typically, SSL multipliers have been used in combination with continuously tunable backward wave oscillator (BWO) tubes as input radiation sources, see for example the experiments in Ref. [3, 4]. However, only a fraction of the fundamental output radiation generated by the BWO is coupled into the SSL multiplier due to the whole coupling setup limitations. An additional downside to this experimental setup is the inability to measure exact value of the electric field inside the SSL. Recently, though, we have obtained good agreement between theory and experiments by estimating the field inside the SSL and adjusting the $\alpha$ parameter to a series of harmonic power measurements using a nonlinear least-squares curve-fitting algorithm based on the Levenberg-Marquardt method [3, 4, 22]. Once $\alpha$ is determined, one can estimate the corresponding value of the electric field amplitude and the effective input field power $P_{in}$ inside the superlattice. In order to determine the exact value of $P_{in}$, we assume that the field delivered by a backward wave oscillator at the SSL is an incoming plane wave of amplitude $E_{ac}$ and frequency $\omega$, and the characteristic impedance of free space is $Z_0 \approx 377\ \Omega$. Thus, the $P_{in}$ is related to $\alpha$ parameter as

$$P_{in}\ (W) = \gamma_a\ \omega^2 \alpha^2, \qquad (7)$$

where $\gamma_a = \hbar^2 S_A/(Z_0 e^2 d^2)$ is a parameter which describes the amount of action performed by the plane wave. For a spot size, $S_A$=112 nm × 2 $\mu$m, given by the surface of the SSL exposed to the field, we obtain $\gamma_a = 75 \times 10^{-15}$ J s. The plane assumption for the optical power is consistent with modelling the input oscillating field as $E_{ac} \cos(\omega t)$.

Now, we can then introduce the intrinsic conversion efficiency $\eta_l$ for the conversion of input field into radiation at the l[th] harmonic

$$\eta_l = \frac{P_l}{P_{in}}. \qquad (8)$$

The intrinsic conversion efficiency used is based on the estimated amplitude of the input field inside the sample and thus independent of geometrical factors that characterize different setups. It is thus a more rigorous and powerful way to model the microscopic frequency multiplication mechanism efficiency.

We further consider room temperature operation. Figure 2 shows the dependence of third-($\eta_3$) and fifth harmonic ($\eta_5$) conversion efficiencies on the parameter α, for input field frequencies ranging from the GHz (black curves in Fig. 2) up to the THz range (blue curves in Fig. 2).

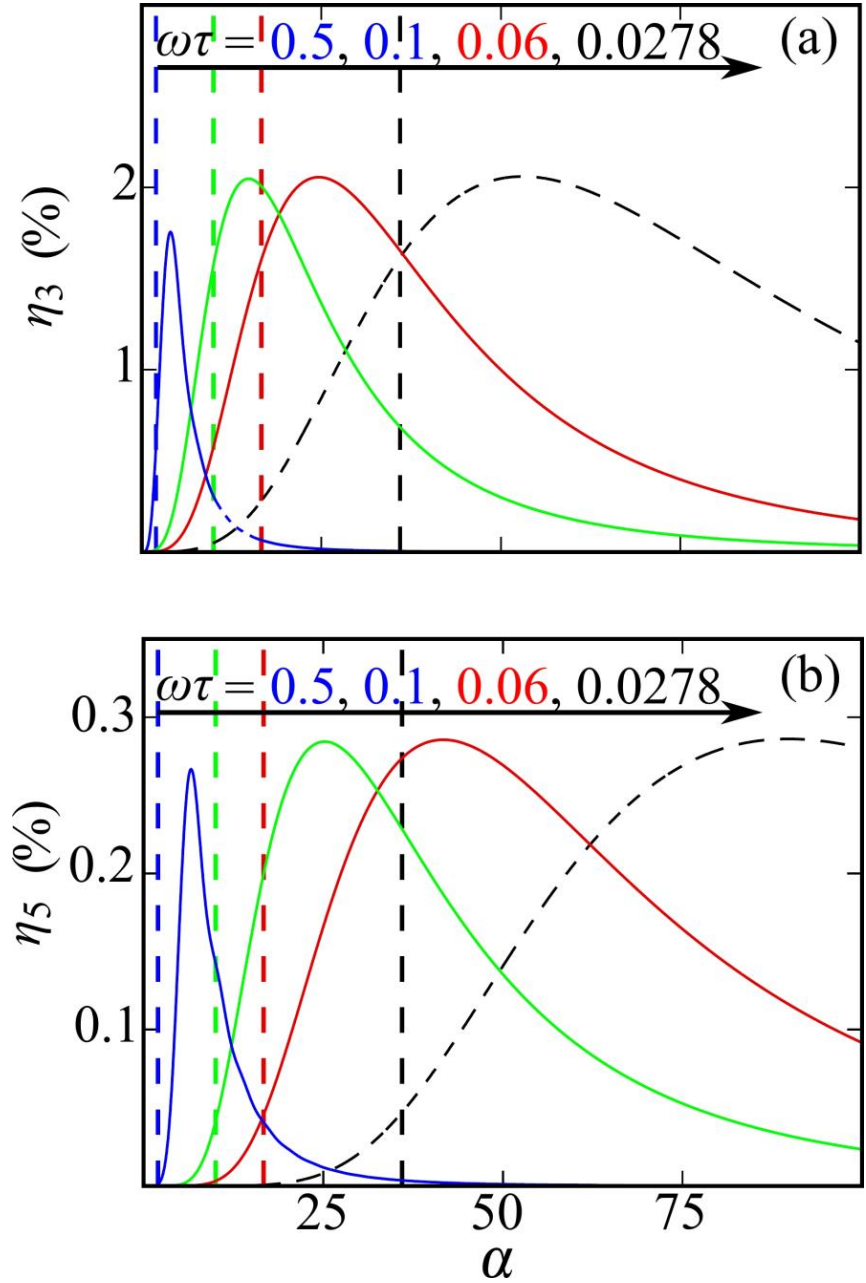

FIG. 2. (a) Intrinsic conversion efficiency as a function of $\alpha = eE_{ac} d/\hbar\omega$ for conversion of input pump signal into the third harmonic (a) $\eta_3$ and the fifth harmonic (b) $\eta_5$. The vertical dashed lines correspond to $\alpha_c$. The colors designate the different frequencies scaled by the relaxation time of the sample, $\tau = \frac{\hbar}{\Gamma} = 31$ fs (calculated by NEGF [3, 4]) for input radiation frequencies $\nu$ =141, 305, 508, 2534 GHz ($\nu = \omega/2\pi$).

Figure 2 shows that by increasing the parameter $\alpha$, the efficiency $\eta_3$ increases roughly up to 2 % slightly after $\alpha$ exceeds $\alpha_c$. Therefore, the irradiation of superlattice with an oscillating input source leads conventionally to frequency tripling due to the nonlinearity of the voltage-current characteristic. The magnitude of this peak is comparable with previous measurements which reported that a highly doped GaAs/AlAs superlattice allowed the conversion of pump to third-harmonic radiation at 300 GHz with an efficiency of approximately 1% [7]. In addition, the values of parameters used to calculate $n_l$ in Fig. 1 reproduce with high accuracy the input parameters of the experiments in [3, 4]. For sufficiently large $\alpha$, well beyond the critical input field strength, the efficiency drops because the generated radiation containing the third harmonic saturates while the input power increases. If the amplitude of the pump does not reach the negative differential conductance, the optimal efficiency for the generation of third harmonic radiation does not surpass 1.7 %. This effect becomes more significant for higher frequencies of the input field as shown in Fig. 2(a). Here, the dominant mechanism for narrowing down the margin of the efficiency can be directly attributed to reduced frequency modulation (small $\alpha$) of current oscillations. For the fifth harmonic, the situation qualitatively remains the same but the maximum radiation efficiency $\eta_5$ is significantly smaller (~0.25 %). Next, we turn to the case in which imperfections in the structure lead to asymmetric scattering processes under forward and reverse bias. Figure 3 shows the efficiency $\eta_3$ as a function of $\alpha$ for different asymmetry coefficients δ when the SSL sample is subjected to an external GHz field.

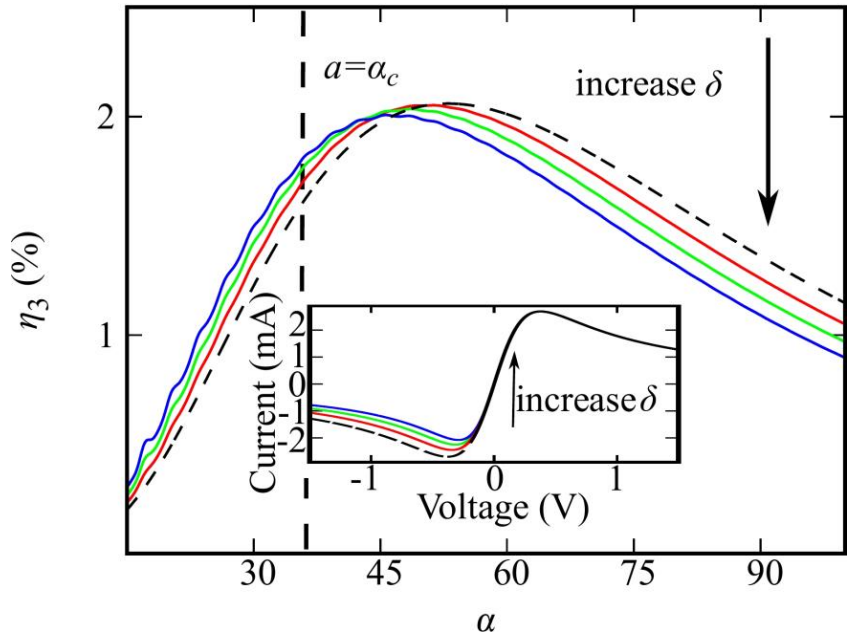

FIG. 3. Dependence of the intrinsic conversion efficiency $\eta_3$ on the parameter $\alpha$ for the conversion of input signal ($\nu$ =141 GHz). The curves below the dashed line (ideal superlattice) have a different asymmetry parameter $\delta = \Gamma^+/\Gamma^-$, which increases by $\delta$ =1.1, 1.2, 1.3. Inset: Current-voltage curves calculated with a variation $\delta$ =1.1, 1.2, 1.3. The vertical line designates the critical field for exciting SSL into states of NDC.

As the asymmetry increases, the peak efficiency decreases and the corresponding maxima are centered at different locations due to the different intraminiband relaxation processes, therefore reducing the maximal efficiency $\eta_3$ (δ=1, dashed curve in Fig. 3). In particular, for the current oscillations in the NDC regime, the maximal oscillator efficiency can be reduced by 0.5 % for a highly asymmetric SSL (δ=1.3, blue curve in Fig. 3). The voltage-current characteristics, which were calculated using Eqs. (1)-(4), are illustrated in the inset of Fig. 3. As was expected, the changes in the dependence of dc current on the static voltage are simultaneously related to the polarity of the applied electric field and the differences of the sample interfaces. It should be pointed out that in the range of values investigated, asymmetry in current voltage can only limit the conversion efficiency of the odd harmonics but not suppress their generation. On the other hand, as expected, breaking the symmetry has the opposite effect on even harmonics, as evident from Fig. 4.

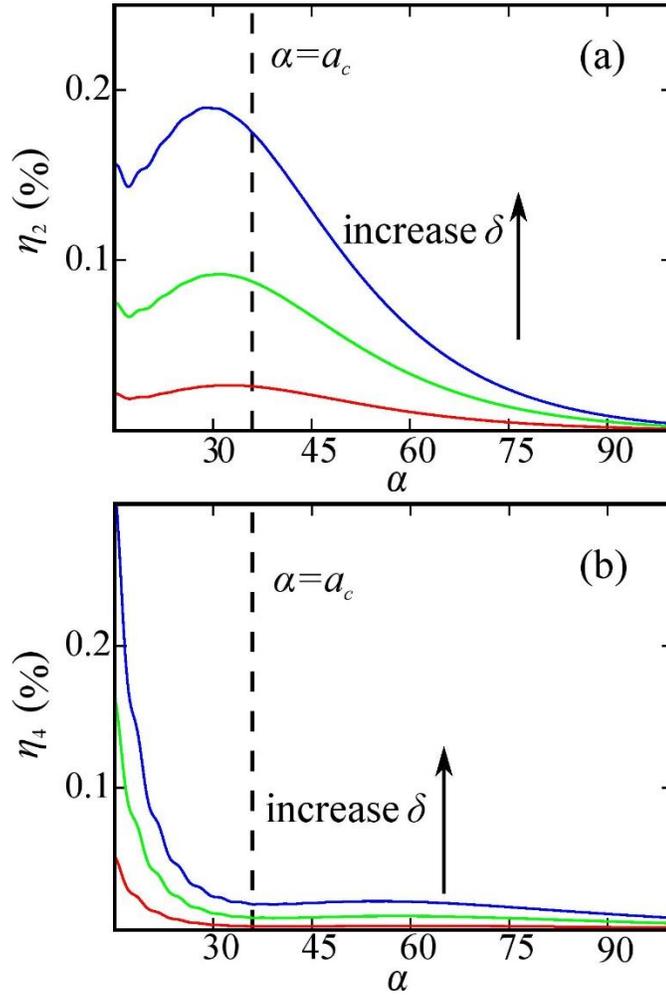

FIG. 4. (a) Intrinsic conversion efficiency $\eta_2$ of the SSL for conversion of input signal ($\nu$=141 GHz, $\omega\tau$=0.0278) for different values of $\alpha$ parameter. (b) Intrinsic conversion efficiency $\eta_4$ of the SSL for conversion of input signal ($\nu$=141 GHz, $\omega\tau$=0.0278) for different values of $\alpha$ parameter. From bottom to the top the asymmetry parameter increases by $\delta$ =1.1, 1.2, 1.3. The vertical lines designate the critical field for exciting SSL into states of NDC.

This picture depicts the generation of the second and fourth harmonics, which is impossible for an ideal superlattice and also reveals notable efficiency for highly asymmetric interfaces. Note that the conversion efficiency for the second harmonic radiation can almost reach the 0.2 % (blue curve in Fig. 4.a). For the fourth harmonic, the optimal efficiency decreases with increasing parameter $a$ (see Fig. 4.b). It is emphasized that the above values for $α$, are based on an ensemble of measurements which represent the most accurate and reliable inputs for detecting harmonic generation. Even though signatures were measured up for a large spectrum of even harmonics, the signal-to-noise ratio was reliable only for $a > 10$ at the detector for a convincing quantitative analysis [3, 4]. Furthermore, even harmonic have been measured and explained consistently by the approach used here, but the waveguide used had a low frequency cut-off that prevented second harmonic detection [3, 4]. At this point, it is important to mention that there are a number of imperfections in real SSLs, among them unwanted charges in the structure. In the NDC range, these can turn into propagating charge domains [6], which strongly influence the overall asymmetry of the problem and play a role in the emission processes [7]. In fact, charge domains should become important for $a > α_c$ and input frequency $ωτ \ll 1$ as the one used to calculate efficiency in Fig. 1 (dashed curve). However, in our analysis we assume that the electric instability due to NDC does not prevent the efficient interaction of the input field with Bloch oscillating electrons. We should note that a domain-free generation is possible if the input field frequency belongs to the high-frequency part of the terahertz range. Alternatively, the suppression of space-charge instabilities holds either for a SSL oscillator with a microwave pump or the special design of a structure that compensates their effects [19, 23]. Future work will take into consideration domain effects using a more complete microscopic NEGF approach.

In summary, using a tested ansatz-solution that can partially include breaks of symmetry in current flow, we have calculated the power conversion efficiency of input signals into harmonics in semiconductor superlattices. The intrinsic conversion efficiency used is based on the estimated amplitude of the input field inside the sample and thus independent of geometrical factors that characterize different setups. We found that deviations from a completely anti-symmetric current-voltage characteristic can lead to radical changes to the output power. Therefore, our study suggests that both unintentional and designed structural variations of superlattices should strongly affect both even and odd order nonlinearities. From the point of view of power generation, special designs of superlattice interfaces may contribute to the efficient coverage of GHz to THz ranges for both even and odd harmonics, expanding the frequency range of superlattice multipliers.

**ACKNOWLEDGEMENTS**

Research supported by the Czech Science Foundation (GAČR) through grant No. 19-03765S.

**APPENDIX**

The asymmetric current in our NEGF calculations stems from the interface roughness self-energy and the parameters used to describe the characteristics of each interface. The interface roughness $\xi_j(\mathbf{r})$ is defined as a fluctuation of the interface $j$ width at position $z_j$, with an order of magnitude about one monolayer. The roughness $\xi_j(\mathbf{r})$ stems from an autocorrelation function, characterized by a Gaussian distribution with height $\eta$ and length $\lambda$

$$\langle \xi_j(\mathbf{r})\xi_j(\mathbf{r}') \rangle = \eta^2 e^{-\frac{|\mathbf{r}-\mathbf{r}'|}{\lambda}} \tag{A1}$$

The corresponding interface potential is

$$V_{ab}^{\text{rough}}(\mathbf{r}) = \sum_j \xi_j(\mathbf{r}) \Delta E_j\, \psi_\alpha^*(z_j)\psi_\beta(z_j), \tag{A2}$$

where $\psi_\alpha(z_j)$ is the wave function of Wannier state α at interface j and $\Delta E_j$ is the intersubband offset. The exact parameters used for the AlAs over the GaAs interface are $\Delta E_j = -1$ eV, $\eta = 0.1$ nm, and λ=5 nm whereas for the GaAs over the AlAs are $\Delta E_j = 1$ eV, $\eta = 0.2$ nm, and λ=5 nm. These are the same parameters used in Refs. [3, 4]. To calculate the interface roughness self-energy which describes the influence of the interface scattering, we employed the second Born approximation [3, 24]. Including the interface roughness self-energy in the complete NEGF calculations, we obtained the current voltage characteristic which allowed us to determine the parameters $j_0^+$, $j_0^-$, i.e. the maximum and minimum current density and corresponding critical energies $U_c^+$ and $U_c^-$. A direct connection between calculated global dephasing and these extrema is given by $\Gamma^+ = U_c^+$ and $\Gamma^- = U_c^-$. Thus, comparison with experiments can give a direct measure of global dephasing/scattering processes in the structure, since critical voltages and energies are the same except for the electron charge connecting V to eV.

Here, we additionally describe how the Green's functions and semiclassical (Boltzmann equation) approaches are connected in the calculations. By employing the balance equation approach [6, 13, 18] to incorporate the effect of interface scattering and the ansatz-solution of the NEGF calculations [3, 4], we express the functions $Y$ and $K$ [see Eq. (2)] in the following form

$$Y(U) = \beta j_0 \frac{2U\tau_{eff}/\hbar}{1 + (U\tau_{eff}/\hbar)^2}, \quad K(U) = \beta \frac{2j_0}{1 + (U\tau_{eff}/\hbar)^2}, \quad (A3)$$

with the effective scattering time $\tau_{eff} = \sqrt{\tau_\varepsilon \tau_v}$ and $\beta = \sqrt{\tau_v/\tau_\varepsilon}$, where $\tau_\varepsilon$ and $1/\tau_v = 1/\tau_\varepsilon + 1/\tau_{int}$ are the phenomenological scattering constants for electron energy and miniband electron velocity, and $\tau_{int}$ is the scattering rate related to the interface scattering processes. This is just the result (2) with the additional factor β reducing (increasing) the maximum (minimum) magnitude of the current $j_0$. Hence, we directly link the effective scattering time $\tau_{eff}$ to the parameter $\Gamma$ calculated by the NEGF calculations. More importantly, the balance equation approach will be modified as compared to the standard equations (3) according to

$$j_0 = \begin{cases} \beta^+ j_0, & U < 0 \\ \beta^- j_0, & U \geq 0 \end{cases}. \quad (A4)$$

Then the ratio between the maximum and minimum values of the current becomes

$$\frac{j_0^+}{j_0^-} = \delta, \quad (A5)$$

where $\delta = \beta^-/\beta^+ = (\tau_{eff}^-/\tau_{eff}^+)$ is the asymmetry coefficient introduced in Eq. (4). One can now consider the effects in terahertz generation for different values of $\tau_{eff}^-$ ($\Gamma^-$) through the parameter $\delta$. To perform these calculations, we benefited from the exact solution of NEGF approach which determines with accuracy the ratio (4).